\documentclass[12pt]{article}

\usepackage[a4paper,left=30mm,right=20mm,top=20mm,bottom=20mm]{geometry}
\usepackage{graphics}
\usepackage{amsmath}
\usepackage{amssymb}
\usepackage{epsfig}
\usepackage{cite}
\usepackage{xcolor}
\usepackage[unicode]{hyperref}
\title{Test of the TMD gluon density in a proton with the longitudinal structure function $F_L(x,Q^2)$}

\author{A.V.~Lipatov$^{1,2}$, G.I.~Lykasov$^{2}$, M.A.~Malyshev$^{1,2}$}

\begin{document}

\maketitle

\begin{center}

{\it $^{1}$Skobeltsyn Institute of Nuclear Physics, Lomonosov Moscow State University, 119991, Moscow, Russia}\\
{\it $^{2}$Joint Institute for Nuclear Research, 141980, Dubna, Moscow region, Russia}\\

\end{center}

\vspace{0.5cm}

\begin{center}

{\bf Abstract }

\end{center}

We investigate the dependence of the deep inelastic proton structure 
function $F_L(x, Q^2)$ on different forms of the
transverse momentum dependent (TMD, or unintegrated)
gluon distribution.
We present a comparison of theoretical results with the
latest experimental data taken by the H1 and ZEUS Collaborations at HERA.
We demonstrate that these data, despite of having quite large uncertainties, 
can test different kinds of the gluon TMD in a proton.
Moreover, different phenomenological models at low $x$ could be tested by future experiments
on deep inelastic scattering.

\indent


\vspace{1cm}


\newpage


It is well known that the differential cross
section of deep inelastic electron-proton scattering (DIS)
can be expressed in terms of 
proton structure functions (SFs) $F_2(x, Q^2)$ and $F_L(x, Q^2)$.
The longitudinal structure function $F_L(x,Q^2)$ is a very sensitive 
characteristic of Quantum Chromodynamics (QCD) since it is directly 
related to the gluon content of the proton.
In fact, it is equal to zero in the naive parton model with spin $1/2$ partons (quarks) and has got 
nonzero values in the framework of perturbative QCD.
Indeed, knowledge about the quark and, especially, the gluon content of the proton
is necessary for any theoretical study of high energy processes performed within the QCD.
Moreover, it is necessary for future experiments on deep inelastic scattering
planned at the Large Hadron electron Collider (LHeC) 
and Future Circular hadron-electron Collider (FCC-he),
where facilities for using electron-proton center-of-mass energies 
$\sqrt s = 1.3$~TeV and $3.5$~TeV are proposed.
Usually, the number of collider data is employed to constrain
the gluon and quark density functions in a proton.
In this sense, an appropriate analysis of available HERA data on the structure function $F_L(x, Q^2)$ taken by 
the H1\cite{FL-H1} and ZEUS\cite{FL-ZEUS} Collaborations at $\sqrt s = 318$~GeV 
is a rather important and interesting task.

However, calculated next-to-leading (NLO) pQCD corrections to the leading order (LO) predictions for
$F_L(x, Q^2)$ are large and negative at small $x$ (see, for example\cite{FL-NLO-Thorne-1, FL-NLO-Thorne-2}), that could 
even lead to negative $F_L(x, Q^2)$ values at small $x$ and low $Q^2$.
The latter demonstrate limitations of the
applicability of perturbative QCD expansion and the necessity of a resummation 
procedure (see\cite{Resummation-1, Resummation-2, Resummation-3, Resummation-4, Resummation-5, Resummation-6}).
Such resummation can be performed
in the framework of so-called
high energy QCD factorization\cite{HighEnergyFactorization} 
(or $k_T$-factorization approach\cite{kt-factorization})
using the Catani-Ciafaloni-Fiorani-Marchesini (CCFM) evolution equation\cite{CCFM} for 
transverse momentum dependent (TMD, or unintegrated) gluon density in a proton.
This equation resumes large logarithmic terms 
proportional to $\alpha_s^n \ln^n 1/x$ and $\alpha_s^n \ln^n 1/(1 - x)$
and therefore valid at both small and large $x$.
The $k_T$-factorization QCD approach 
supplemented with the CCFM gluon evolution dynamics
is becoming an essential tool 
which allows one to make theoretical predictions for processes
studied at modern and future colliders. More information
can be found in review\cite{TMD-review}.

At present, a special interest is connected with investigation of the TMD gluon
density in a proton. In fact, while a great amount of knowledge about the conventional parton 
density functions (PDFs), obeying Dokshitzer-Gribov-Lipatov-Altarelli-Parisi (DGLAP) equations\cite{DGLAP} has been  
accumulated over past years, the TMD gluon and quark densities are still
poorly known quantities.
Similar to DGLAP-based PDFs,
the TMDs are usually parameterized in a general form and then 
fitted to some experimental data.
So, the latest precision HERA measurements of the DIS structure functions $F_2(x, Q^2)$ and $F_2^c(x, Q^2)$
were used\cite{JH2013} to determine the CCFM-evolved gluon densities in a proton,
where the empirical expression for TMD gluon density at some starting scale $\mu_0^2$ (which is of the 
order of the hadron scale)
with factorized Gauss smearing in transverse momentum was
applied\footnote{There are other popular approaches to evaluate the TMDs, for
example, the Kimber-Martin-Ryskin (KMR) prescription\cite{KMR-LO, KMR-NLO} or the 
Parton Branching (PB) approach\cite{PB1, PB2}.}.
In contrast, in our previous study\cite{LLM-2022}
the initial TMD gluon distribution was chosen 
 in the unfactorised form, which depends on $x$ and
$k_T$ simultaneously. 
This starting gluon density LLM'2022, like the GBW gluon one \cite{GBW1,GBW2}, results in the saturation 
of the dipole cross section at low $Q^2$.  Moreover,  the form of the LLM'2022 input was chosen in order to describe 
the LHC data on the charged hadron production at low transverse momenta $p_T \sim 1$~GeV
within the modified soft quark-gluon string model (QGSM)
\cite{ModifiedSoftQuarkGluonStringModel-1, ModifiedSoftQuarkGluonStringModel-2}.  
Then, the CCFM equation was applied to extend the proposed TMD gluon density (LLM'2022)
in the whole kinematical region.
Several parameters important at moderate and large $x$
have been fitted from the LHC data on inclusive $b$-jet and Higgs boson 
production as well as the latest HERA data on proton structure functions
$F_2^c(x, Q^2)$, $F_2^b(x, Q^2)$ and reduced cross sections $\sigma_{\rm red}^c(x, Q^2)$
and $\sigma_{\rm red}^b(x, Q^2)$, thus moving forward to the global fit 
of TMD gluon density from the collider data.
However, the measurements\cite{FL-H1, FL-ZEUS} of the longitudinal structure function
$F_L(x, Q^2)$ performed by the H1 and ZEUS Collaborations at HERA
were not taken into account, since they have quite large 
experimental uncertainties as compared to other data.
The main goal of our present study is to test 
the LLM'2022 gluon density with available data
on $F_L(x, Q^2)$ in a wide region of $x$ and $Q^2$.

\begin{figure}
\begin{center}
\includegraphics[width=5cm]{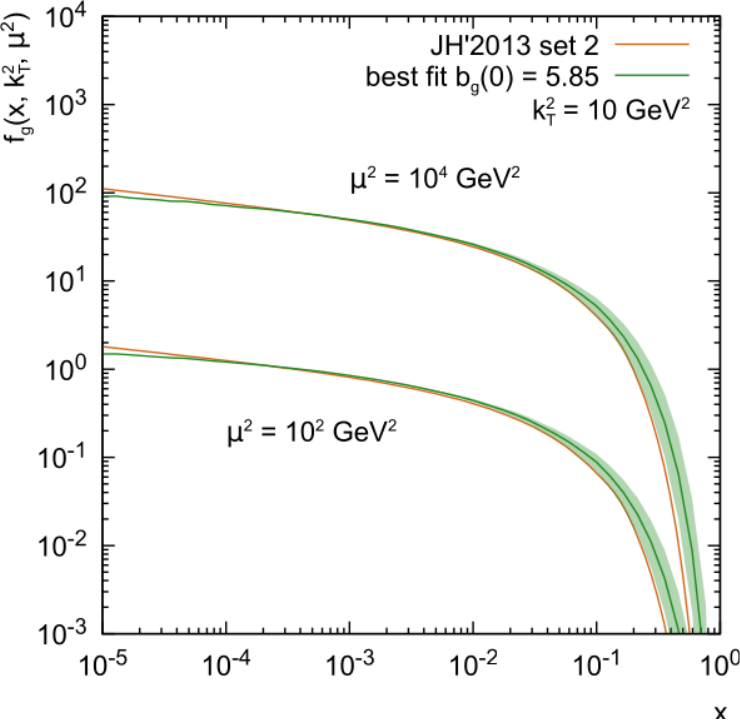}
\hspace*{2cm}
\includegraphics[width=5cm]{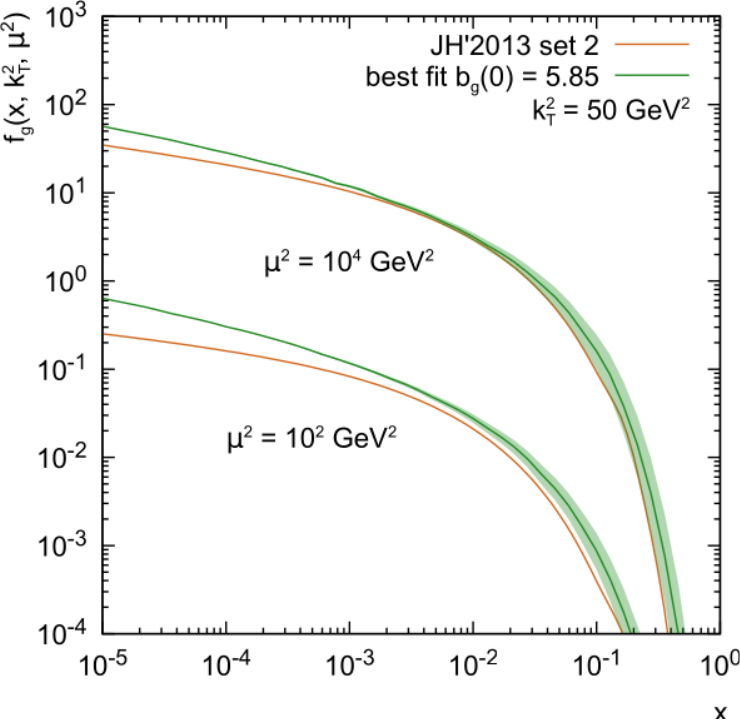}
\includegraphics[width=5cm]{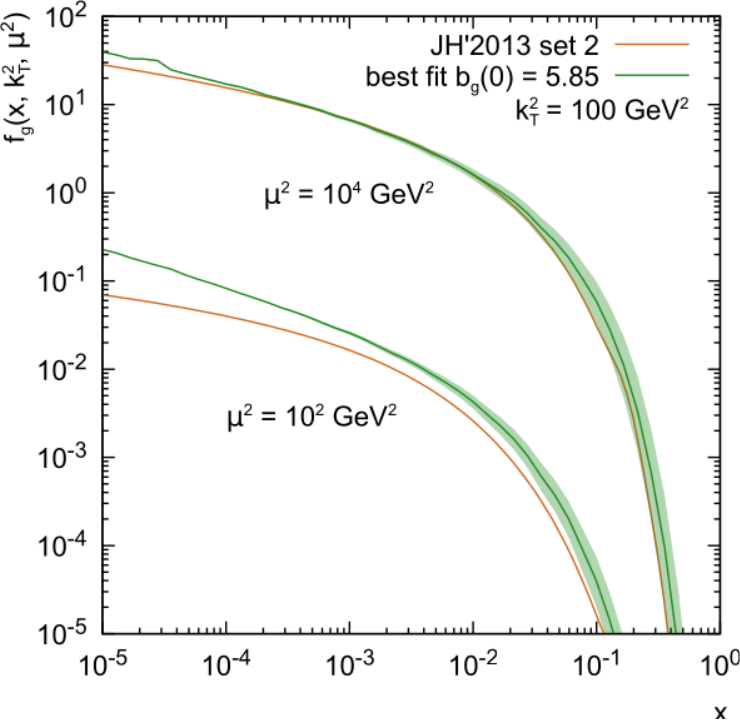}
\hspace*{2cm}
\includegraphics[width=5cm]{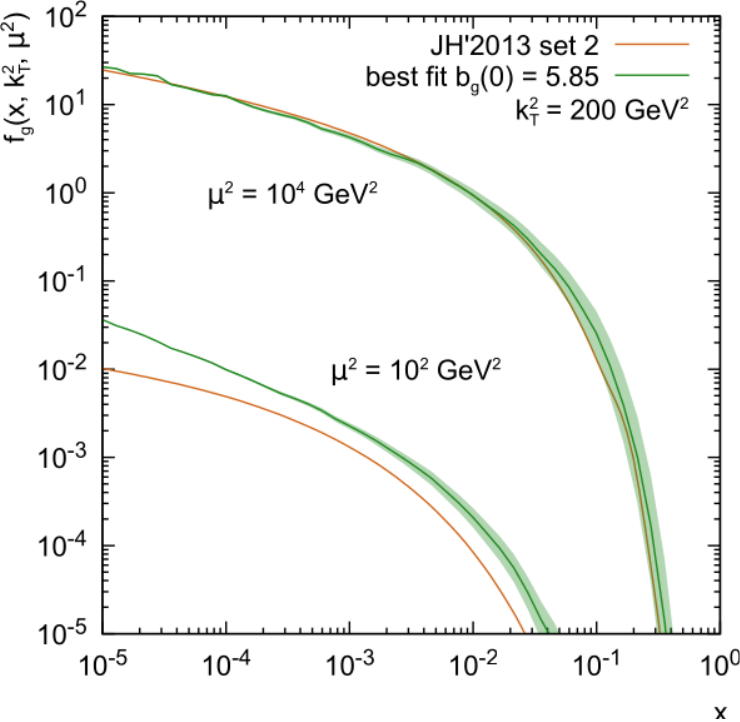}
\caption{The TMD gluon densities in the proton $f_g(x, {\mathbf k}_T^2, \mu^2)$ 
calculated as a function of the longitudinal 
momentum fraction $x$ at different values of the transverse momentum ${\mathbf k}_T^2$ and hard scale $\mu^2$.
Shaded bands represent the uncertainties of the $b_g(0)$ fitting procedure. 
Note that the gluon densities calculated at $\mu^2 = 10^4$~GeV$^2$ are multiplied by a factor of $100$.}
\label{fig:Gluonx}
\end{center}
\end{figure}

\begin{figure}
\begin{center}
\includegraphics[width=5cm]{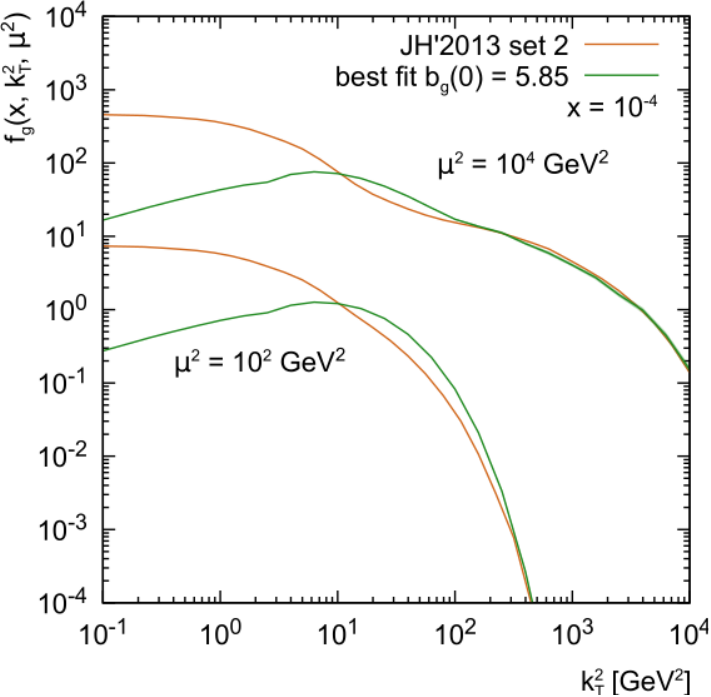}
\hspace*{2cm}
\includegraphics[width=5cm]{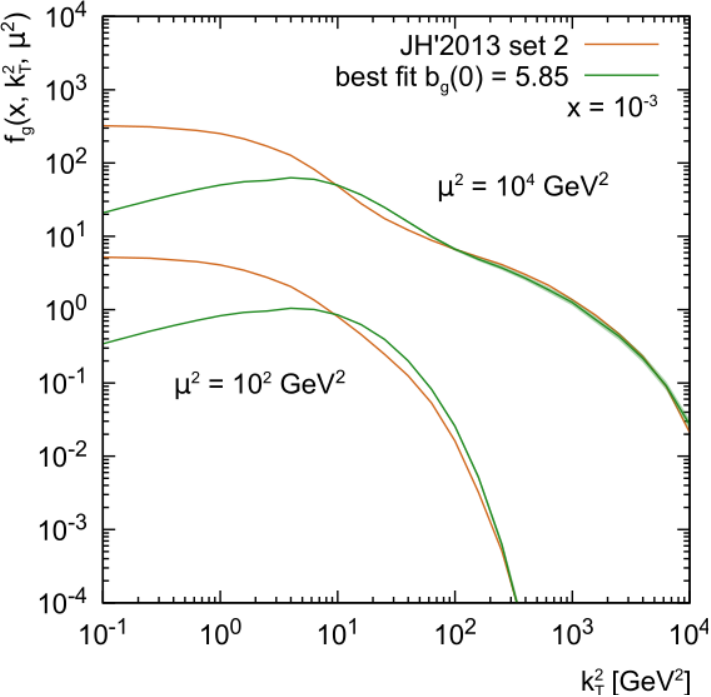}
\includegraphics[width=5cm]{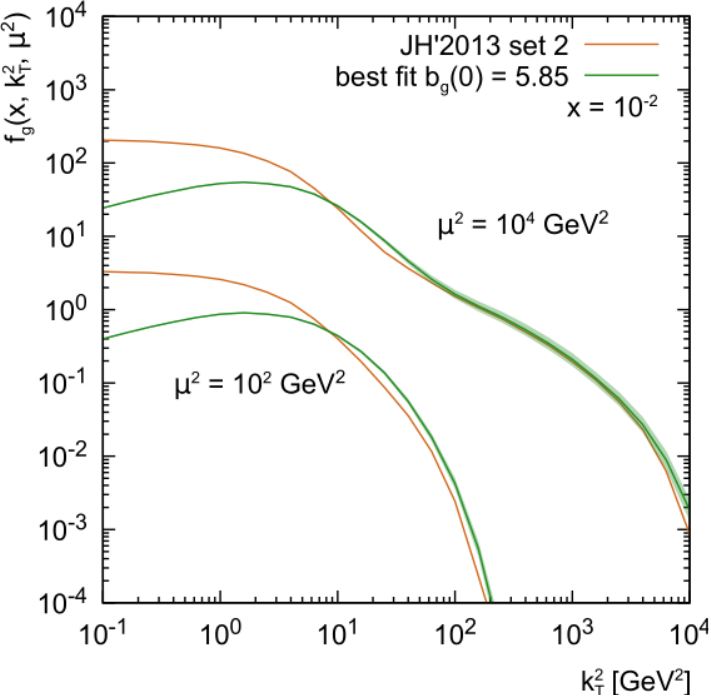}
\hspace*{2cm}
\includegraphics[width=5cm]{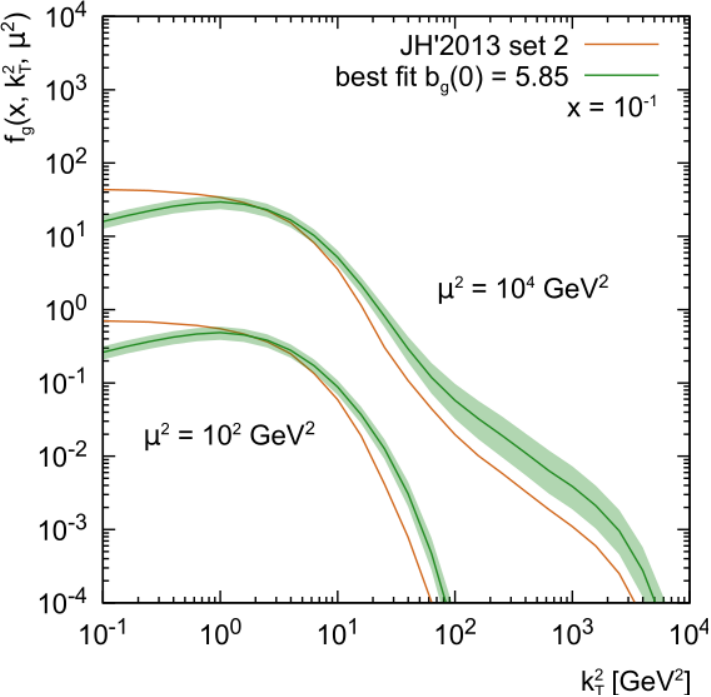}
\caption{The TMD gluon densities in a proton $f_g(x, {\mathbf k}_T^2, \mu^2)$ 
calculated as a function of the transverse momentum ${\mathbf k}_T^2$ at 
different values of the longitudinal 
momentum fraction $x$ and hard scale $\mu^2$.
The notations are the same as in Fig.~\ref{fig:Gluonx}.}
\label{fig:Gluonkt}
\end{center}
\end{figure}

Let us briefly recall our main steps. We started
from quite a general expression for the initial TMD gluon distribution\cite{LLM-2022}:
\begin{gather}
  f_g(x, {\mathbf k}_T^2) = c_g (1-x)^{b_g} \sum_{n = 1}^3 \left(c_n R_0(x) |{\mathbf k}_{T}|\right)^n e^{-R_0(x)|{\mathbf k}_{T}|},
  \label{eg:OurGluon1}
\end{gather}
\noindent
where
\begin{gather}
  R_0(x) = {1\over Q_0} \left( {x\over x_0} \right)^{\lambda/2}, \quad b_g = b_g(0) + {4 C_A \over \beta_0} \ln {\alpha_s(Q_0^2) \over \alpha_s({\mathbf k}_T^2)},
  \label{eg:OurGluon2}
\end{gather}
\noindent
with $C_A = N_C$, $\beta_0 = 11 - 2N_f/3$, $x_0 = 4.1 \cdot 10^{-5}$ and 
$\lambda = 0.22$.
From the best description of recent LHC data on charged 
soft hadron production in $pp$ collisions at low $p_T$ and the mid-rapidity region
(collected at $\sqrt s = 0.9$, $2.36$, $7$ and $13$~TeV) within the 
modified QGSM the parameters 
$c_1 = 5$, $c_2 = 3$, $c_3 = 2$ and $Q_0 = 1.233$~GeV, essential
at small $x$, have been extracted. The 
values $b_g(0)=5.854$ and $c_g = 0.173$ were determined
from the simultaneous fit of HERA and LHC data on 
some hard QCD processes performed
in the framework of $k_T$-factorization approach,
as it was mentioned above.
All the details of calculations and fitting procedure can be found in ref.\cite{LLM-2022}.
Here we only note that a rather good value $\chi^2/d.o.f. = 2.2$ is achieved.
The obtained TMD gluon density in a proton is shown in Figs.~\ref{fig:Gluonx} and \ref{fig:Gluonkt} 
as a function of the proton longitudinal momentum fraction $x$ and
gluon transverse momentum ${\mathbf k}_T^2$ for several values of hard scale $\mu^2$. 
The shaded bands represent uncertainties of the fitting procedure,
which become important at large $x \geq 10^{-1}$.
For comparison, we also show the CCFM-evolved TMD
gluon distribution JH'2013 set 2\cite{JH2013}, which is often 
used in phenomenological applications.
In contrast with our approach, the $x$-dependence of JH'2013 set 2
input has quite a general form proportional to $x^a (1 - x)^b$
with parameters derived from the high-precision HERA data on the proton structure functions
$F_2(x, Q^2)$ and $F_2^c(x, Q^2)$ at $x < 5 \cdot 10^{-3}$ and $Q^2 > 3.5$~GeV$^2$.
One can see that both TMD gluon densities have a remarkably different $x$
and ${\mathbf k}_T^2$ behaviour, especially in the region of small ${\mathbf k}_T^2$.
As a consequence, the LLM'2022 gluon density provides a 
better description of the considered hard QCD processes (see\cite{LLM-2022}).
Moreover, it saturates at a scale lower than the ones predicted by the popular 
dipole Golec-Biernat-W\"usthoff (GBW) model\cite{GBW1, GBW2}.
Currently, the LLM'2022 gluon density is implemented into the Monte-Carlo 
event generator \textsc{pegasus}\cite{PEGASUS} and included in the \textsc{tmdlib} package\cite{TMDLib2}, 
which is used by the \textsc{cascade}\cite{CASCADE} and \textsc{KaTie}\cite{KATIE} Monte-Carlo generators.

In this paper we focus on investigation of the longitudinal SF $F_L(x,Q^2)$,
which is related to the DIS differential cross section as follows:
\begin{gather}
\frac{d\sigma}{dxdy}=\frac{2\pi\alpha^2}{xQ^4}\left[\left(1-y+\frac{y^2}{2}\right)F_2(x,Q^2)-\frac{y^2}{2}F_L(x,Q^2)\right],
\end{gather}
\noindent 
where $Q^2$ is the virtuality of the probing photon, $x$ is the Bjorken variable and $y$ is the inelasticity. 
Our calculations below are based on the formulas\cite{SFs-our} and here we 
strictly follow our previous consideration in all aspects.
So, according to the $k_T$-factorization prescription, the proton structure function $F_L(x, Q^2)$
can be calculated as a convolution
\begin{gather}
F_L(x,Q^2)=\int\limits_x^1\frac{dz}{z}\int d{\mathbf k}_T^2\sum e_f^2\hat C^g_L(x/z,Q^2,m_f^2,{\mathbf k}_T^2)f_g(z,{\mathbf k}_T^2, \mu^2),
\end{gather}
\noindent
where the summation is performed over flavors $f$ of quarks with 
electric charges $e_f$ and masses $m_f$. 
The hard coefficient function $C^g_L(x/z,Q^2,m_f^2,{\mathbf k}_T^2)$ corresponds to the quark-box
diagram for off-shell (dependent on the incoming gluon virtuality) 
photon-gluon fusion subprocess and was calculated earlier\cite{SFs-our}.
Numerically, we set the charm and beauty masses 
to $m_c = 1.67$~GeV and $m_b = 4.75$~GeV and use the massless 
limit to evaluate the corresponding contributions from the light quarks.
Also, we apply the $2$-loop formula for the strong coupling constant $\alpha_s$
with $N_f = 4$ quark flavours at $\Lambda_{\rm QCD} = 200$~MeV.

\begin{figure}
\begin{center}
\includegraphics[width=15cm]{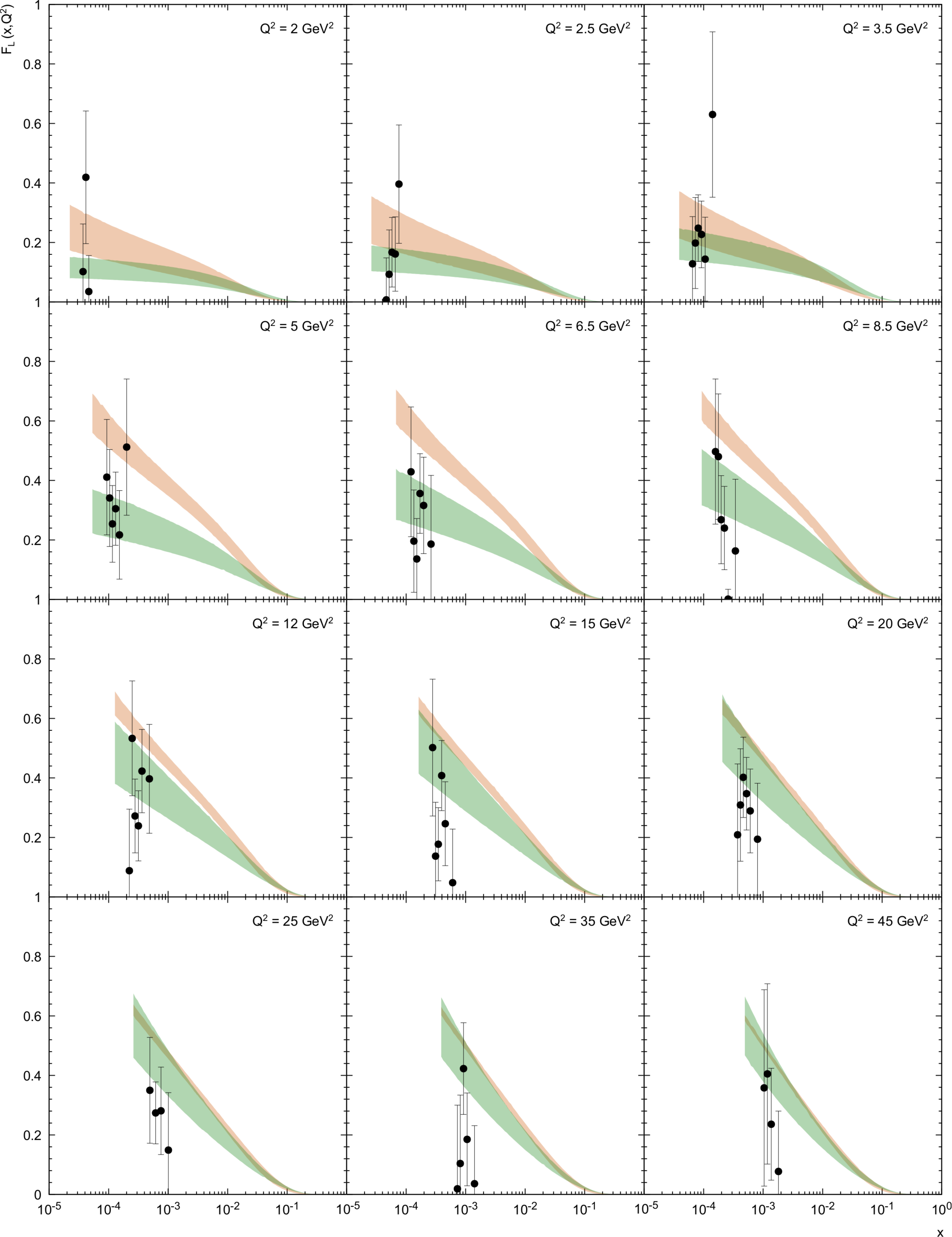}
\caption{
The longitudinal SF $F_2(x,Q^2)$ calculated at different $Q^2$. The green 
band correspond to the result obtained with the LLM'2022 gluon density 
with scale uncertainties, while the red one represents results calculated with the JH'2013 set 2 one. 
The experimental data are from H1\cite{FL-H1}.}
\label{fig:H1}
\end{center}
\end{figure}

\begin{figure}
\begin{center}
\includegraphics[width=15cm]{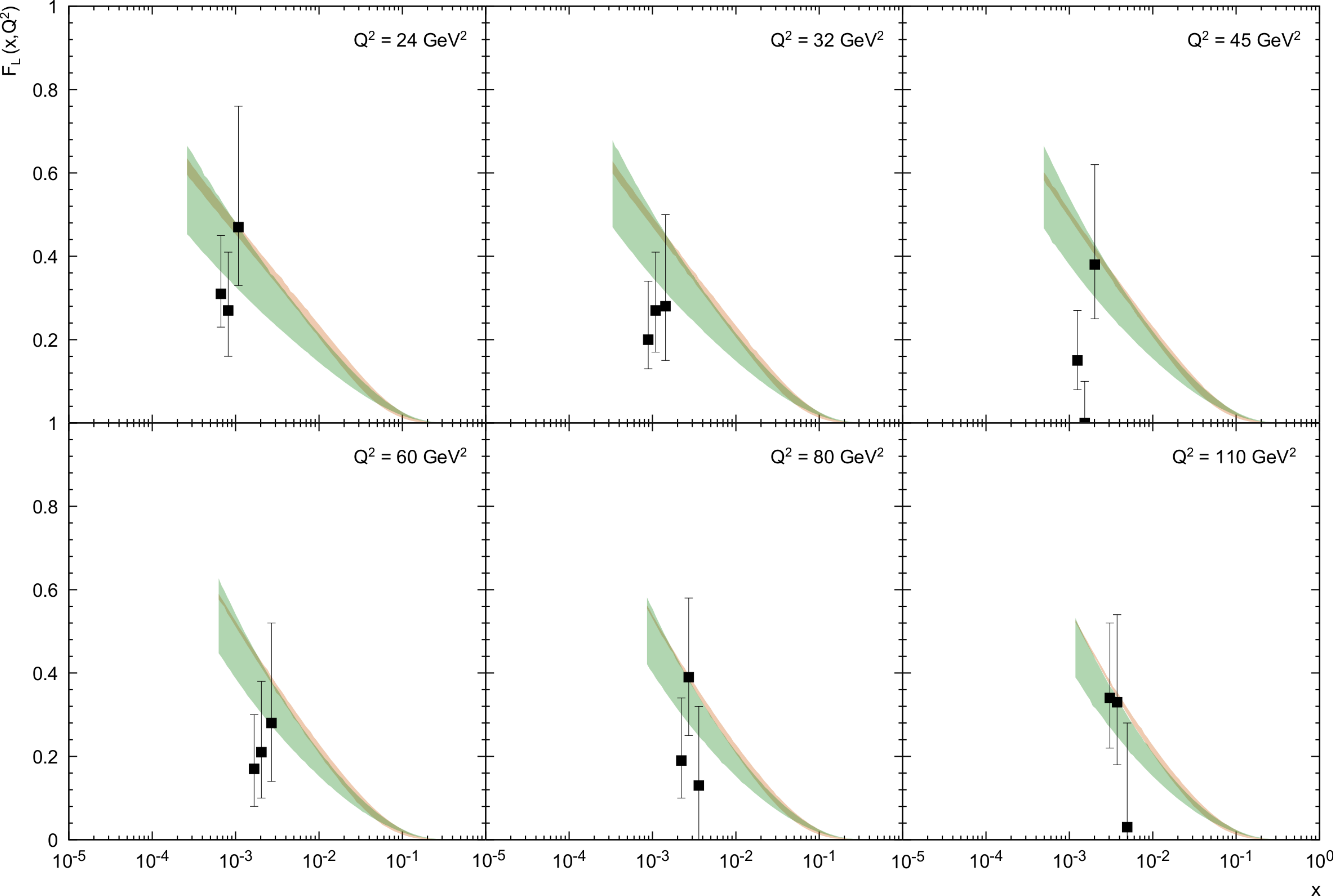}
\caption{
The longitudinal SF $F_2(x,Q^2)$ calculated at different $Q^2$. The notations are the same as in Fig.~\ref{fig:H1}. 
The experimental data are from ZEUS\cite{FL-ZEUS}.}
\label{fig:ZEUS}
\end{center}
\end{figure}

The results of our calculations are shown in Figs.~\ref{fig:H1} and \ref{fig:ZEUS}. Here we 
plot the evaluated $F_L(x, Q^2)$ as a function of $x$ for different values of $Q^2$
in comparison with available HERA data, taken by the H1\cite{FL-H1} and ZEUS\cite{FL-ZEUS} Collaborations. 
In addition, we also show the results obtained with the JH'2013 set 2 gluon 
density. 
The shaded bands correspond to theoretical uncertainties of our calculations 
connected with the choice of hard scales. Note that in the case of the 
JH'2013 set 2 gluon density we use auxiliary '$+$' and '$-$' distributions 
corresponding to the variated renormalization scales (see\cite{JH2013}).
So, we find that the LLM'2022 gluon density
describes the HERA data quite well 
within the estimated theoretical and experimental uncertainties.
It tends to slightly overestimate the H1 measurements\cite{FL-H1} at large $Q^2$, 
but the predictions are still compatible with the data at $\sim2\sigma$. 
We obtained, also, that the JH'2013 set 2 gluon density provides 
a bit worse description of the HERA data, especially at low $Q^2$.
The better agreement achieved with the LLM'2022 distribution 
is an immediate consequence of using the physically motivated 
expression (\ref{eg:OurGluon1}) for the starting TMD gluon density.
In fact, at low $Q^2$ the relatively small gluon ${\mathbf k}_T^2$
are probed, where the difference between 
the considered TMD gluon distributions becomes large, as it is shown in
Fig.~\ref{fig:Gluonkt}.
This big difference can be due to the saturation of the dipole cross section at low $Q^2$
\cite{GBW1,GBW2}, which is taken into account by the LLM'2022 gluon.
So, our calculations demonstrate
that available experimental data for the proton structure function $F_L(x, Q^2)$ 
are sensitive to the TMD gluon densities, and the relatively low $Q^2$ region 
could help to clearly distinguish the latter.

To conclude, we have tested the recently proposed TMD gluon density 
in a proton (LLM'2022) with the longitudinal structure function $F_L(x,Q^2)$ in 
a wide region of $x$ and $Q^2$. 
We have compared corresponding predictions with the
experimental data taken by the H1 and ZEUS Collaborations at HERA 
as well as with results obtained with another CCFM-based gluon distribution, JH'2013 set 2. 
We have found that the predictions with the LLM'2022 gluon density do not contradict available
data in a wide $Q^2$ range.
Moreover, they show a better description of the data than the JH'2013 set 2 distribution
due to more physically motivated expression for the initial TMD gluon density.
Thus, the measurements of $F_L(x,Q^2)$ could verify different 
TMDs, especially at not very large $Q^2$, where the further effects of QCD evolution do not play an essential role.
It could be important for experiments at future 
colliders, such as LHeC or FCC-he, and for 
improvement of different small-$x$ phenomenological models.

{\sl Acknowledgements}. We thank S.P.~Baranov, A.V.~Kotikov, H.~Jung, S.~Taheri Monfared 
for their important comments and remarks. We also thank G.~Pontecorvo
for careful reading of our manuscript.
Studies related with derivation of the LLM'2022 gluon density were supported by the 
Russian Science Foundation under grant~22-22-00387. 
All other performed calculations were supported by the Russian Science Foundation, grant~22-22-00119.

\bibliography{FL}

\end{document}